\documentclass[twocolumn,10pt,english]{article}
\usepackage{babel}
\usepackage{graphicx}
\usepackage{amsmath}
\usepackage{bm}
\usepackage{latexsym}
\begin{document}
\title{\textbf{Casimir force acting on bodies in media:\\Why textbooks are wrong}}
%
\author{\underline{Dirk-Gunnar Welsch} and Christian Raabe\\[8pt]
\normalsize Theoretisch-Physikalisches Institut,
Friedrich-Schiller-Universit\"{a}t Jena,\\[4pt]
\normalsize Max-Wien-Platz 1, 07743 Jena, Germany\\[4pt]
\normalsize Corresponding author: Dirk-Gunnar Welsch (D.-G.Welsch@tpi.uni-jena.de)
}
%
\date{}
\maketitle
\thispagestyle{empty}
The problem of appropriately defining energy and/or momentum of
the macroscopic electromagnetic field in linear, causal media 
has been a subject of controversial discussions for a long time. 
In this context, approximations like quasimonochromatic fields 
or lossless media have typically been employed to derive
local balance equations from the phenomenological Maxwell equations.
Although the meaning of the emerging expressions
often has not been clear at all, they have been taken for granted 
in the study of QED effects associated with media, without 
worrying about their limitations. A typical example, which has even 
entered textbooks, has been the incorrect extension
of the well-known Lifshitz formula for the Casimir force 
between two planar dielectric bodies separated by vacuum to 
the case where the interspace is not empty but filled with a medium.

To answer the question of the effect of a medium,
we first present general expressions for the
Casimir force acting on linearly responding magnetodielectric
bodies, which are not necessarily placed in vacuum but may 
also be surrounded by a linear magnetodielectric medium, basing 
our calculations on exact QED in linear, causal media \cite{ref2}.
Material dispersion and absorption are fully taken into account.
In our approach, the Lorentz force acting on the internal charges and
currents of the medium is regarded as the basic quantity, 
with the internal charges and currents being expressed in terms 
of the medium polarization and magnetization.
It is shown that the result is in full agreement with
microscopic oscillator models, which are widely used for 
the description of linear media. In particular, the result reveals
that the use of Minkowski's stress tensor in the calculation 
of Casimir forces---a method employed in the literature---is wrong
even for a nonabsorbing medium.
%

\begin{figure}[htb]
\centering
\includegraphics[width=7.5cm]{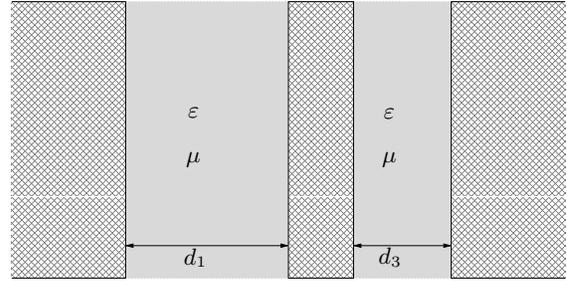}
\caption{\label{fig1}
Casimir plate embedded in a nonempty planar cavity.}
\end{figure}

We have applied the theory to planar structures, with special
emphasis on two multilayered plates the interspace between 
which is filled with a medium,
and on a homogeneous plate embedded in a planar cavity 
filled with a medium (see Fig.~\ref{fig1}). The generalized Lifshitz 
formula obtained in the former case shows that---in contrast 
to the belief in the literature---the Casimir stress \emph{depends}
on the position within the interspace in general. In the latter case,
standard approximations (such as the assumption of almost 
perfect reflection) lead to a force (per unit area) of
\begin{displaymath}
F = \frac{\hbar c \pi^2}{240}\,
\sqrt{\frac{\mu}{\varepsilon}}\,
\left(\frac{2}{3}+\frac{1}{3\varepsilon\mu}
\right)
\left(
\frac{1}{d_3^4}
-\frac{1}{d_1^4}\right),
\end{displaymath}
which for $d_{1(3)}$ $\!\to$ $\!\infty$ is the generalization 
of Casimir's well-known original formula. Comparison with the force formula
\begin{displaymath}
\label{L18}
F^\mathrm{(M)}
=\frac{\hbar c \pi^2}{240}\,
\sqrt{\frac{\mu}{\varepsilon}}
\left(
\frac{1}{d_3^{4}}
-\frac{1}{d_1^{4}}\right),
\end{displaymath}
which may be found on the basis of Minkowski's stress tensor and
has erroneously been predicted (for \mbox{$\mu$ $\!=$ $\!1$}) 
in the literature, clearly shows that if the plate is embedded in a medium
(of static permittivity $\varepsilon$ and permeability $\mu$),
then the correct force can noticeably differ from
the erroneously predicted one.

\end{document}